\documentstyle[11pt,newpasp,twoside]{article} \def\edcomment#1{\iffalse\marginpar{\raggedright\sl#1\/}\else\relax\fi}
\reversemarginpar 

\begin{document} \title{Theory of star formation in the central
kiloparsec} 

\author{Bruce G. Elmegreen} \affil{IBM Research Division, T.J. Watson
Research Center, PO Box 218, Yorktown Hts., NY 10598 USA} 

\begin{abstract} Star formation by gravitational instabilities,
sequential triggering, and turbulence triggering are briefly reviewed in
order to compare the various mechanisms that are observed in main galaxy
disks with those in the inner kiloparsec regions. Although very little
is known about inner galaxy triggering mechanisms, there appear to be
examples that parallel the well-known mechanisms around us. The balance
may shift slightly, however, because of the high densities in nuclear
regions. The sequential triggering mode, in which one generation of
stars triggers another, should be less prominent compared to
gravitational instabilities and turbulence triggering when the ambient
density is so high that the local dynamical time scale is less than the
lifetime of an O-type star. Regardless of this shift in balance, the
star formation rate seems to follow the same column density scaling law
as in main disks, probably because it is everywhere saturated to the
maximum possible value allowed by the fractal structure of the interstellar medium.
\end{abstract} 

\section{Introduction} 

There are essentially three distinct star formation mechanisms that
include most of the observations: gravitational instabilities leading to
dense clouds and then stars by fragmentation, sequential triggering of
dense clouds by other stars, and turbulent triggering of dense clouds by
supersonic compression in a turbulent flow. These processes operate in
the main disks of galaxies. We consider here whether they also operate
in the inner kpc regions where the conditions are sufficiently different
that the proportion of the three may change. 

\subsection{Gravitational Instabilities} 

Gravitational instabilities in the main disks of galaxies produce
structures on kpc scales, including flocculent and grand-design spiral
arms, and regularly spaced star-forming regions and other giant cloud
complexes inside spiral arms (e.g., M51: Kuno et al. 1995). Secondary
gravitational instabilities may occur on smaller scales in the cooled
and colliding parts of these primary instabilities (e.g., Wada \& Norman
1999, 2001). 

Nuclear disks have similar structures. Some galaxies have nuclear grand
design spiral arms (e.g., D15 in Coma -- Caldwell, Rose, \& Dendy 1999;
NGC 5248 -- Laine et al. 1999)
and others have long irregular arms that resemble multiple-arm density
waves in main disks (e.g., NGC 7469 and VIIZw031 observed by Scoville et
al. 2000). Massive patches of star formation can form in either of these
arms. Nuclear regions can also have rings with regularly spaced star
formation, as in ESO 565-11 (Buta, Crocker \& Byrd 1999). 

The spatial
scales for nuclear spirals and star formation are much smaller than they
are in main galaxy disks. In nuclear regions, spirals can be only
several hundred pc long and tens of pc wide; even the largest star-forming 
regions are only several tens of pc in size. The smaller sizes
in nuclear regions are the result of a smaller Jeans length, which comes
from a higher ambient density without a significantly higher
velocity dispersion. 

Gravitational instabilities are also clearly present in some larger
starbursts involving whole galaxy disks. NGC 6090 is a merger system
(Mazzarella \& Boroson 1993) with a molecular surface density of
$\sim3\times10^3$ M$_\odot$ pc$^{-2}$ that peaks in a region between the
two nuclei (Bryant \& Scoville 1999). Color maps of this system in
Dinshaw et al. (1999) show four equally-spaced knots of star
formation along an arc with a length of several kpc. These knots are
like the beads in normal spirals arms, but in NGC 6090, they are much
denser. NGC 253 also has an enormous cluster in the central region with
several thousand O-type stars and an age comparable to 1 My (Watson et al. 1996). 
NGC 5253
has a similar dense cluster near the center (Turner, Beck, \& Ho 2000).
These giant clusters do not appear to be simple extrapolations of a
smooth cluster mass function because there is nothing else like them in
the galaxies. They are also not near obvious pressure sources or shells
as if they were sequentially triggered, nor are they part of a fractal
hierarchy of structures as might result from turbulence. They appear to
be isolated cases of gravitational collapse to single giant clusters. 

\subsection{Sequential Triggering} 

Many of the clusters in the solar neighborhood have been triggered by
external pressures from existing HII regions and stellar winds. Reviews
of this process are in Elmegreen (1998) and Elmegreen et al. (2000). 
Examples in the
recent literature include the Rosette nebula (Phelps \& Lada 1997),
Orion nebula (Bally et al. 1987; Lada et al. 1991; Dutrey, et al. 1991;
Reipurth, Rodriguez \& Chini 1999), Trifid nebula (Lefloch \& Cernicharo
2000), rho Ophiuchus (de Geus 1992), Upper Sco OB association (Preibisch
\& Zinnecker 1999), and W5 (Kerton \& Martin 2000). 

Occasionally a supernova remnant is seen to have triggered star
formation along the periphery. This should be rare because supernova
remnants are usually short-lived compared to the triggering time.
G349.7+0.2 seems to be an example (Reynoso \& Mangum 2000). Other shells
have triggered star formation (Xie \& Goldsmith 1994; Yamaguchi et al.
2001), but they are older than single supernova remnants and were
probably formed by a combination of supernova and stellar wind pressures
(McCray \& Kafatos 1987). 

Giant shells in other galaxies have triggered star formation too (Brinks
\& Bajaja 1986; Puche et al. 1992; Wilcots \& Miller 1998; Steward et
al. 2000; Stewart \& Walter 2000). IC 2574 (Walter \& Brinks 1999;
Steward \& Walter 2000) has HII regions on the perimeter of a shell with
clusters $\sim3$ My old and another cluster in the center of the shell
that is 11 My old. 

There is also evidence for sequential triggering in starburst regions,
like 30 Dor (Walborn et al. 1999), but not much evidence yet for this
process in the central kpc regions of galaxies outside the Milky Way. In
our own central region, the giant molecular cloud Sgr B2 was recently
found to contain shells in NH$_3$ maps, and these shells contain hot
cores on their periphery -- evidence for triggering of massive stars
(Martin-Pintado et al. 1999). There is little clear evidence yet for
shells and two-stage star formation in other galaxy nuclei.
Alonso-Herrero, Ryder, \& Knapen (2001) suggested that HII regions in
the nuclear region of NGC 2903 were excited by stars triggered by older
clusters seen nearby as IR sources, but the clusters inside the HII
regions have not been found yet, and they could, in principle, be
excited by radiation escaping from the existing IR clusters. 

Clear evidence for sequential triggering in the nuclear regions of
galaxies may be difficult to get. The high rates of shear and the small
disk scale heights in inner galaxy regions should distort any shells
that form, making them hard to recognize. The length scales for
triggering are usually small too, perhaps 10 pc in a local GMC and 1 pc
in a dense nuclear GMC. This makes the resolution required very high. 

Sequential triggering could actually be less important than
spontaneous gravitational instabilities in inner galaxy disks because
the dynamical time, $\left(G\rho\right)^{-1/2}$, at the extreme density of
a nuclear disk is comparable to or less than the stellar evolution time
of $\sim1$ My. This means that spontaneous instabilities could fill a
region with star formation before the dying stars from the first
generation get a chance to supernova and trigger much themselves. 

\subsection{Turbulence Triggering} 

Turbulence triggers cloud and star formation by compressing gas that
cools and becomes gravitationally unstable. In main galaxy disks, this
compression can produce star fields with a fractal character (Gomez et
al. 1993; Elmegreen \& Elmegreen 2001) and it can give a correlation
between age and scale (Harris \& Zaritsky 1999) or between duration of
star formation and scale (Efremov \& Elmegreen 1998; Battinelli \&
Efremov 1999; Scalo \& Chappell 1999; Nomura \& Kamaya 2001). Star
formation times can be very short when clouds are formed by supersonic
motions (Ballesteros-Paredes, Hartmann, \& Vazquez-Semadeni 1999). 

There have been many numerical simulations of cloud and star formation
triggered by compressible turbulence (MacLow \& Ossenkopf 2000; Pichardo
et al. 2000; Klessen, Heitsch, \& MacLow 2000; V\'azquez-Semadeni,
Gazol, \& Scalo 2000; Semelin \& Combes 2000; Ostriker, Stone, \& Gammie
2001; Heitsch, Mac Low, \& Klessen 2001; Toomre \& Kalnajs 1991; Wada \&
Norman 1999, 2001). The converging regions increase the density and
produce clouds, which can then become gravitationally unstable and make
clusters and stars. 

The clumpy and fractal structure of supersonically turbulent gas helps
explain the mass functions of clouds and clusters (Fleck 1996; Elmegreen \&
Falgarone 1996; St\"utzki et al. 1998; Elmegreen \& Efremov 1997; Elmegreen
2001a). This mass function, which is $\sim M^{-2}dM$, is found at the
high mass end of old globular clusters (Ashman, Conti, \& Zepf 1995),
for young globular clusters in starburst regions (Whitmore \& Schweizer
1995; Zhang \& Fall 1999), open clusters in the solar neighborhood
(Battinelli et al. 1994), and OB associations in galaxies (Kennicutt,
Edgar, \& Hodge 1989; Elmegreen \& Salzer 1999; McKee \& Williams 1997).

The case for turbulence triggering in the central kpc is not well
documented. There is no observation yet of a size-age or size-duration
correlation for star formation, nor is there good evidence for fractal
structure in the positions of young stars. Nevertheless, the dust
spirals in some nuclear regions have a fractal quality reminiscent of
turbulence (Elmegreen, Elmegreen, \& Eberwein 2001), and the morphology
of these spirals suggests shear is very important. Thus the spatial
correlations and space-time correlations that show up well in 
low-shear environments such as the Large Magellanic Cloud may be present but
difficult to recognize in the central kpc regions of spiral galaxies. 

\section{Star formation rates} 

The three processes of star formation mentioned above combine to give
the net star formation rate in a galaxy. This total rate may be
saturated to the maximum possible value given the structure of the gas;
i.e., saturated to the value at which every suitable region forms stars on a
dynamical time scale that is proportional to the inverse square root of
the density (Elmegreen 2001b). The mechanism of star formation may not
matter for the total rate; if one mechanism does not work well, then
another will take over to give the same total rate. 

The combination of processes also means that while gravity, star
formation pressures, and turbulence may structure the positions and
densities of gas clouds, the actual triggering processes inside these
clouds may be a different combination of mechanisms. A cloud may
form by one process and get star formation triggered by another. Thus it
is often difficult to say how a particular star formed. 

There are several correlations that should be expected for star
formation that operates in a turbulent gas on a dynamical time scale.
First, the mass of the largest cluster should increase with the square
of the number of clusters (as observed by 
Whitmore 2000) because of the size-of-sample
effect. Thus regions of active star formation should have not only more
clusters but also bigger clusters. 

Second, the relative fraction of star formation that occurs in the form
of clusters should increase with the total star formation rate (Larsen
\& Richtler 2000). This is because the total pressure scales
approximately with the square of the gas column density in a galaxy,
$P_{ISM}\propto \Sigma_{gas}^2$, and the star formation rate scales with
this column density approximately as ${\rm SFR}\propto
\Sigma_{gas}^{1.4}$ (Kennicutt 1998). Thus $P_{ISM}\propto {\rm
SFR}^{1.4}$. In addition, for virialized clusters, the
velocity dispersion is $c^2\sim0.2GM/R$ and the internal pressure is
$P_{int}\sim0.1GM^2/R^4$. These two equations give a relation between
mass and pressure:
\begin{equation}M\sim6\times10^3\left(P_{int}/10^8\;{\rm
K\;cm^{-3}}\right)^{3/2} \left(n/10^5\;{\rm
cm}^{-3}\right)^{-2}.\end{equation} The normalization for this relation
comes from the properties of the molecular core near the Trapezium
cluster in Orion (Lada, Evans \& Falgarone 1997). This mass is the
maximum cluster mass since smaller clusters can fragment out. From this
we get $M\propto P_{int}^{3/2}$ for internal pressure $P_{int}$. If the
internal pressure scales with the ambient pressure, then the maximum
cluster mass $M_{max}\propto {\rm SFR}^2$. 

The Larsen \& Richtler (2000) correlation follows from this expression
for maximum mass. The total mass of young dense clusters equals the
integral of the cluster mass weighted by the mass spectrum from the
smallest mass, $M_{min}$, to the maximum mass, $M_{max}$. For an
$n(M)dM=n_0 M^{-2}dM$ spectrum, the normalization factor $n_0$ comes
from the integral of $n(M)dM$ from $M_{max}$ to infinity:
$\int_{M_{max}}^{\infty}n_0M^{-2}dM=1$, which gives $n_0=M_{max}$ and
$n(M)dM=M_{max}M^{-2}dM$. Thus the total cluster mass is $M_{tot}\propto
M_{max}\ln\left(M_{max}/M_{min}\right)$. The log term is slowly varying,
so we get $M_{tot}\propto M_{max}\propto {\rm SFR}^2$. From this, the
fraction of the star formation in the form of dense clusters,
$M_{tot}/{\rm SFR}$, scales directly with the star formation rate:
\begin{equation} M_{tot}/{\rm SFR}\propto {\rm SFR}.\end{equation} This
is essentially the relation found by Larsen \& Richtler (2000). 

A turbulent medium has a distribution of density that is approximately
log-normal (V\'azquez-Semadeni 1994; Nordlund \& Padoan 1999, Klessen
2000; Wada \& Norman 2001). Stars form in the densest gas, and this
accounts for only a small fraction of the total mass. The rest is at a
lower density and is generally faster moving. In the full ISM there are
also several temperatures for the gas, and, although much of it is
turbulent, only the cool component forms stars. The formation of single
or binary stars requires a sufficiently high density in a small region
that starlight is excluded, magnetic diffusion is relatively rapid,
gravity is relatively strong, and the turbulent motions are subsonic and
incompressible. The star formation rate should be approximately the
total mass exceeding this density multiplied by the dynamical rate and
the efficiency for the collapse process itself. Collapse efficiencies
may be several tens of percent once the star formation process starts
(Matzner \& McKee 2000).

\section{Summary}

The various processes of star formation that are observed in main galaxy
disks are also observed in the inner kiloparsec regions where the
densities and star formation rates per unit area are much larger. The
relative proportion of these processes should change, however, as the
ratio of the ambient dynamical time, which is
$\sim\left(G\rho\right)^{-1/2}$ for density $\rho$, to the stellar
evolution time, which is independent of $\rho$, decreases. For high
densities, spontaneous gravitational instabilities at the ambient Jeans
length and the more local instabilities that occur in
turbulence-compressed gas should operate much faster than supernova and
stellar wind triggering, which seem to dominate the onset of star
formation locally. Not much is known about these relative rates,
however. High shear, high extinction, small angular scales, and short
time scales make the determination of specific triggers for star
formation difficult to determine in the inner kiloparsec.

\end{document}